\documentclass[twocolumn,english,aps,prl,floatfix,superscriptaddress,showpacs,amsmath,amssymb]{revtex4}
\usepackage{amsmath}
\usepackage{amssymb}
\usepackage{stackrel}
\usepackage{graphicx}
\usepackage{esint}

\makeatletter
\@ifundefined{textcolor}{}
{%
 \definecolor{BLACK}{gray}{0}
 \definecolor{WHITE}{gray}{1}
 \definecolor{RED}{rgb}{1,0,0}
 \definecolor{GREEN}{rgb}{0,1,0}
 \definecolor{BLUE}{rgb}{0,0,1}
 \definecolor{CYAN}{cmyk}{1,0,0,0}
 \definecolor{MAGENTA}{cmyk}{0,1,0,0}
 \definecolor{YELLOW}{cmyk}{0,0,1,0}
}

\newcommand{\rmc}{{\rm c}}
\newcommand{\rme}{{\rm e}}
\newcommand{\rmd}{{\rm d}}
\newcommand{\rmi}{{\rm i}}

\@ifundefined{textcolor}{}{%
 \definecolor{BLACK}{gray}{0}
 \definecolor{WHITE}{gray}{1}
 \definecolor{RED}{rgb}{1,0,0}
 \definecolor{GREEN}{rgb}{0,1,0}
 \definecolor{BLUE}{rgb}{0,0,1}
 \definecolor{CYAN}{cmyk}{1,0,0,0}
 \definecolor{MAGENTA}{cmyk}{0,1,0,0}
 \definecolor{YELLOW}{cmyk}{0,0,1,0}
}

\usepackage{babel}

\usepackage{babel}

\makeatother

\usepackage{babel}
\begin{document}

\title{Driven-dissipative Ising model: mean-field solution}

\author{G. Goldstein}

\affiliation{Department of Physics, Rutgers University, Piscataway, New Jersey
08854, USA}

\author{C. Aron}

\affiliation{Department of Electrical Engineering, Princeton University, Princeton,
New Jersey 08544, USA}

\affiliation{Instituut voor Theoretische Fysica, KU Leuven, Belgium}

\author{C. Chamon}

\affiliation{Department of Physics, Boston University, Boston, Massachusetts 02215,
USA}
\begin{abstract}
We study the fate of the Ising model and its universal properties
when driven by a rapid periodic drive and weakly coupled to a bath
at equilibrium. The far-from-equilibrium steady-state regime of the
system is accessed by means of a Floquet mean-field approach. We show
that, depending on the details of the bath, the drive can strongly
renormalize the critical temperature to higher temperatures, modify
the critical exponents, or even change the nature of the phase transition
from second to first order after the emergence of a tricritical point.
Moreover, by judiciously selecting the frequency of the field and
by engineering the spectrum of the bath, one can drive a ferromagnetic
Hamiltonian to an antiferromagnetically ordered phase and \textit{vice-versa}. 
\end{abstract}
\maketitle
The Ising model is undoubtedly the most studied model of statistical
mechanics. Besides its equilibrium properties, its coarsening dynamics
following a temperature quench from the paramagnetic to the ordered
phase is also quite well understood~\cite{BrayReview,Exact2d}, even
in the presence of weak disorder~\cite{Dziarmaga,SAALFC,ACLFCP}.
Taking into account the dissipative mechanisms due to the inevitable
coupling of the spin system to an environment has been successful
in the description of important many-body phenomena such as the decay
of metastable phases~\cite{key-30,key-31,key-32,key-33,key-34,key-35},
hysteretic responses~\cite{key-36,key-37,key-38} and magnetization
switching in mesoscale ferromagnets~\cite{key-39,key-40}. As it
is becoming clear these days that driven-dissipative physics, \textit{i.e.}
the balancing of non-equilibrium conditions and dissipative mechanisms,
is a promising route to achieve a new type of control over matter,
a burning question arises: can the Ising model be driven to non-equilibrium
steady states (NESS) with enhanced or even novel properties?

This question has been approached in the context of slowly oscillating
drives (magnetic fields or electrochemical potentials) by means of
Monte-Carlo simulations~\cite{key-35b,key-36,key-37,key-38,key-47,key-48,key-49},
mean-field treatment~\cite{key-41,key-42,key-43,key-44,key-45,key-46},
or other analytical techniques~\cite{key-50,key-51,key-52,key-53}.
One of the key results is the existence of a so-called dynamical phase
transition, where the cycle-averaged magnetization becomes non-zero
in a singular fashion. This has recently been supported by experimental
evidence in the dynamics of thin ferromagnetic films~\cite{key-54}.

In this Letter, we focus on the Ising model driven by a rapidly oscillating
magnetic field $h\cos(\omega t)$. We depart from the usual Floquet
engineering approach to many-body phases, mostly directed towards
cold-atomic systems~\cite{PolkovnikovReview}, by including a dissipative
mechanism, namely by weakly coupling the system to an external equilibrium
bath. The properties of the latter are kept generic in order to study
the influence of its spectrum on the dynamics. Dissipation is the
natural counterpart of driving and our results are an unquestionable
proof that the presence of a bath can have far-reaching consequences.

We access the non-equilibrium steady states by means of a Floquet
mean-field approach. We derive the mean-field self-consistent equation
for the magnetization and use it to derive the non-equilibrium phase
diagram. Whenever analytical solutions are beyond reach, we complete
the picture with numerical results. Our main results are to show how
to combine drive (\textit{i.e.} $h$ and $\omega$) and dissipation
($\textit{i.e.}$ mostly the low-energy spectrum of the bath) to increase
the critical temperature $T_{\rmc}$, to modify the critical exponent
$\beta_{T}$, as well as to change the order of the phase transition.
Additionally, we show that the drive can, in the presence of carefully
selected baths, convert a ferromagnetically ordered system to an antiferromagnetic
order, and \textit{vice versa}.

\begin{figure}[t]
\includegraphics[width=7cm]{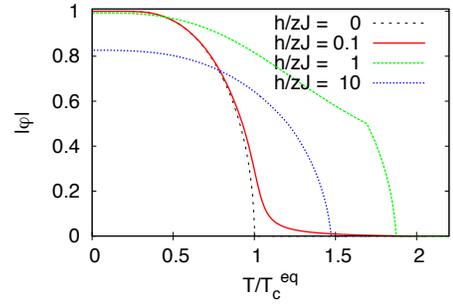} \protect\caption{{\footnotesize{}(color online) Mean-field magnetization $|\varphi|$
as a function of the temperature $T$ of the bath (super-Ohmic, $s=1$)
for different values of the drive $h$ given in the key: $h=0$ (equilibrium),
$h\ll\omega$, $h\sim\omega$ and $h\gg\omega$. The critical temperature
for $h\ll\omega$, $T_{\rmc}=(1+s)T_{\rmc}^{{\rm eq}}$, is computed
exactly in Eq.~(\ref{eq:Tc2}). In the temperature range $T_{\rmc}^{{\rm eq}}<T<T_{\rmc}$,
$|\varphi|\sim(h/\omega)^{2/s}\omega/zJ$. ($\omega=zJ=T_{\rmc}^{{\rm eq}}$).}}
\end{figure}

\paragraph{Model.}

The total Hamiltonian is composed of the system, the bath and the
system-bath Hamiltonians, $H(t)=H_{S}\left(t\right)+H_{B}+H_{SB}$
with (we set $\hbar=k_{{\rm B}}=1$) \begin{subequations}\label{eq:H}
\begin{align}
H_{S}\left(t\right)= & -J\sum_{\langle ij\rangle}\sigma_{i}^{z}\sigma_{j}^{z}-h\cos\left(\omega t\right)\sum_{i}\sigma_{i}^{z}\;,\label{eq:System_hamiltonian}\\
{H}_{B}= & \sum_{i,\alpha}\omega_{\alpha}\; b_{i,\alpha}^{\dagger}\, b_{i,\alpha}\;,\label{eq:Bath_hamiltonian}\\
{H}_{SB}= & \sum_{i,\alpha}t_{\alpha}\,\sigma_{i}^{x}\left(b_{i,\alpha}+b_{i,\alpha}^{\dagger}\right)\;.\label{eq:System_Bath_hamiltonian}
\end{align}
\end{subequations} The spin $1/2$'s, represented at each site $i$
of the bipartite lattice by the usual Pauli operators $\sigma_{i}^{x,y,z}$,
are interacting through a nearest-neighbor interaction $J$. $h$
is the strength of the periodic drive with frequency $\omega\equiv2\pi/\tau$
(we choose $\omega\geq0$). Equilibrium conditions are recovered for
$h=0$ or $\omega=0$. Notice that in the absence of an environment,
the drive has trivial consequences on the dynamics of the Ising model.
Indeed, as $[\sigma_{i}^{z},H(t)]=0$ at all times, all the degrees
of freedom are conserved quantities. The environment is composed of
local baths expressed in terms of a collection of non-interacting
bosonic modes labelled by $\alpha$, with energy $\omega_{\alpha}$
and with creation and annihilation operators given by $b_{i,\alpha}^{\dagger}$
and $b_{i,\alpha}$. Each is in equilibrium at temperature $T\equiv1/\beta$
and we assume it is a ``good bath'', \textit{i.e.} it has a very
large number of degrees of freedom and it remains in thermal equilibrium.
Below, we replace $\sum_{\alpha}$ by $\int\rmd\epsilon\,\rho(\epsilon)$
where $\rho(\epsilon)$ is the bath density of state. Without loss
of generality, the chemical potential is set to 0 and $\rho(\epsilon<0)=0$.
$H_{{\rm SB}}$ is responsible for thermal spin flips and the parameters
$t_{\alpha}$ control the strength of the spin-bath couplings. After
integrating out the bath degrees of freedom, the bath will enter the
reduced problem \textit{via} the hybridization function $\nu(\epsilon)\equiv|t({\epsilon})|^{2}\rho\left(\epsilon\right)$.
The low-energy behavior of the hybridization $\nu(\epsilon)\stackrel[0<\epsilon\to0]{}{\sim}\epsilon^{1+s}$
characterizes whether the bath is Ohmic ($s=0$), sub-Ohmic ($s<0$),
or super-Ohmic ($s>0$). Note that we do not consider additional system-bath
coupling terms such as $\sigma_{i}^{y,z}\,(b_{i,\alpha}+b_{i,\alpha}^{\dagger})$
because they do not induce any qualitative change in the non-equilibrium
dynamics.

\paragraph{Floquet mean-field description.}

The time-dependent mean-field Hamiltonian corresponding to $H(t)$
in Eq.~(\ref{eq:H}) is the one of a single spin coupled to its local
bath, and reads $\bar{H}(t)=\bar{H}_{S}\left(t\right)+\bar{H}_{B}+\bar{H}_{SB}$
with \begin{subequations} 
\begin{align}
\bar{H}_{S}\left(t\right)= & -zJ\varphi\left(t\right)\sigma^{z}-h\sigma^{z}\cos\left(\omega t\right)\;,\label{eq:System_hamiltonian-1}\\
\bar{H}_{B}= & \sum_{\alpha}\omega_{\alpha}\; b_{\alpha}^{\dagger}\, b_{\alpha}\;,\label{eq:Bath_hamiltonian-1}\\
\bar{H}_{SB}= & \sum_{\alpha}t_{\alpha}\,\sigma^{x}\left(b_{\alpha}+b_{\alpha}^{\dagger}\right)\;.\label{eq:System_Bath_hamiltonian-1}
\end{align}
\end{subequations} Here, $\varphi(t)$ is the expectation value of
$\sigma^{z}(t)$ which serves as the order parameter, and $z$ is
the coordination number of the bipartite lattice. When the coupling
the bath is weak (see the discussion below), the spin subsystem can
be seen as quasi-isolated during many periods of the drive. There,
the Floquet theorem states that the instantaneous eigenstates of the
time-periodic Hamiltonian $\bar{H}_{{\rm S}}(t)$ can be written in
the form $|\psi_{\alpha}(t)\rangle=\rme^{-\rmi E_{\alpha}t}|\psi_{\alpha}^{{\rm P}}(t)\rangle$
where $E_{\alpha}$ is a so-called Floquet quasi-energy and $|\psi_{\alpha}^{{\rm P}}(t)\rangle$
is periodic: $|\psi_{\alpha}^{{\rm P}}(t+\tau)\rangle=|\psi_{\alpha}^{{\rm P}}(t)\rangle$.
Owing to the fact that $\sigma^{z}$ is a conserved quantity, we may
choose our Floquet eigenstates to simultaneously diagonalize $\sigma^{z}$.
Note that this also implies that $\varphi(t)$ is a constant (at least
between two events induced by the weakly-coupled bath). Altogether,
the instantaneous eigenstates of $\bar{H}_{{\rm S}}(t)$ are simply
given by 
\begin{align}
\left|\uparrow\!\left(t\right)\right\rangle = & \,\rme^{+\rmi\left[zJ\varphi\, t+\frac{h}{\omega}\sin\left(\omega t\right)\right]}\;\left|\uparrow\right\rangle %
%=
%\rme^{+\rmizJ\varphi\, t}\,\sum_{n}J_{-n}\left({h}/{\omega}\right)
%\,\rme^{\rmin\omegat}
%\;\left|\uparrow\right\rangle %
=\rme^{-\rmi\epsilon_{\uparrow}t}\;\left|\uparrow^{{\rm P}}\!(t)\right\rangle \;,\label{eq:Floquet_up}\\
\left|\downarrow\!\left(t\right)\right\rangle = & \,\rme^{-\rmi\left[zJ\varphi\, t+\frac{h}{\omega}\sin\left(\omega t\right)\right]}\;\left|\downarrow\right\rangle %
%=
%\rme^{-\rmizJ\varphi\, t}\,\sum_{n}J_{+n}\left({h}/{\omega}\right)
%\,\rme^{\rmin\omegat}
%\;\left|\downarrow\right\rangle %
=\rme^{-\rmi\epsilon_{\downarrow}t}\;\left|\downarrow^{{\rm P}}\!(t)\right\rangle \;,\label{eq:Floquet_down}
\end{align}
from which one identifies the Floquet quasi-energies and the periodic
states, reading 
\begin{align}
\epsilon_{\uparrow}\equiv-zJ\varphi\;,\quad & \left|\uparrow^{{\rm P}}\!(t)\right\rangle =\sum_{n}J_{n}\left({h}/{\omega}\right)\rme^{-\rmi n\omega t}\left|\uparrow\right\rangle \;,\\
\epsilon_{\downarrow}\equiv zJ\varphi\;,\quad & \left|\downarrow^{{\rm P}}\!(t)\right\rangle =\sum_{n}J_{n}\left({h}/{\omega}\right)\rme^{+\rmi n\omega t}\left|\downarrow\right\rangle \;,
\end{align}
where $J_{n}$ are the Bessel functions of the first kind.

\paragraph{Rates.}

The bath is inducing incoherent transitions between the eigenstates.
The transition rate $R_{\uparrow\downarrow}$ from $\left|\uparrow\right\rangle $
to $\left|\downarrow\right\rangle $ can be obtained by means of a
Floquet-Fermi golden rule~\cite{key-28}: 
\begin{align}
R_{\uparrow\downarrow}(\varphi)=2\pi\sum_{m\in\mathbb{Z}}|A_{\uparrow\downarrow}^{m}|^{2}\; g\left(\epsilon_{\uparrow}-\epsilon_{\downarrow}+m\omega\right)\;,\label{eq:Rates}
\end{align}
with $g\left(\epsilon\right)\equiv\nu(\epsilon)[1+n_{{\rm B}}\left(\epsilon\right)]+\nu(-\epsilon)n_{{\rm B}}(-\epsilon)$
where the Bose-Einstein distribution $n_{{\rm B}}(\epsilon)\equiv1/(\rme^{\beta\epsilon}-1)$
and 
\begin{align}
A_{\uparrow\downarrow}^{m} & \equiv\int_{0}^{\tau}\!\!\frac{\rmd t}{\tau}\left\langle \downarrow^{P}\!\!(t)\right|\sigma^{x}\left|\uparrow^{P}\!\!(t)\right\rangle \rme^{\rmi m\omega t}=J_{m}\left({2h}/{\omega}\right).\label{eq:A_m}
\end{align}
A similar expression can be obtained for $R_{\downarrow\uparrow}(\varphi)$
with $A_{\uparrow\downarrow}^{m}=A_{\downarrow\uparrow}^{-m}$. Note
that the integration over the degrees of freedom of the bath also
contributes to a small renormalization of the spin Hamiltonian (so-called
Lamb-shift) that we neglect.

\paragraph{Steady-state population.}

We stress that the previous analysis is valid only in the case the
bath is \emph{weakly} coupled to the system, \textit{i.e.} the rate
at which it induces spin flips is much smaller than the frequency
of the drive: $R_{\uparrow\downarrow},R_{\downarrow\uparrow}\ll\omega$.
Under these conditions, $\varphi(t)$ is indeed constant over many
periods of the drive and a time-translational invariant non-equilibrium
steady state can establish. Once it is reached, the probabilities
of being in the $|\uparrow\rangle$ and $|\downarrow\rangle$ states
are simply given by 
\begin{align}
P_{\uparrow}^{{\rm NESS}}=\frac{1}{1+R_{\uparrow\downarrow}/R_{\downarrow\uparrow}}\mbox{ and }P_{\downarrow}^{{\rm NESS}}=1-P_{\uparrow}^{{\rm NESS}}\;.\label{eq:Probablities}
\end{align}

\paragraph{Self-consistency condition.}

The probabilities in Eq.~(\ref{eq:Probablities}) allow to compute
the steady-state average magnetization as $|\varphi|=|P_{\uparrow}^{{\rm NESS}}-P_{\downarrow}^{{\rm NESS}}|$.
Therefore, we obtain the self-consistency condition for the mean-field
order parameter 
\begin{equation}
\pm\;\varphi=\frac{R_{\downarrow\uparrow}(\varphi)-R_{\uparrow\downarrow}(\varphi)}{R_{\downarrow\uparrow}(\varphi)+R_{\uparrow\downarrow}(\varphi)}\;.\label{eq:Self_consistency}
\end{equation}
Here, the $+$ sign corresponds to a ferromagnetic order while the
$-$ sign corresponds to an antiferromagnetic order. Making use of
the expression for the rates given in Eq.~(\ref{eq:Rates}), we obtain
\begin{widetext} \begin{subequations} 
\begin{eqnarray}
R_{\downarrow\uparrow}(\varphi)-R_{\uparrow\downarrow}(\varphi)= &  & 2\pi\left|J_{0}\left({2h}/{\omega}\right)\right|^{2}\nu\left(\left|2zJ\varphi\right|\right)\;{\rm sgn}(J\varphi)+2\pi\sum_{n>0}\sum_{a,b=\pm}\left|J_{n}\left({2h}/{\omega}\right)\right|^{2}\, b\,\nu\left(an\omega+2bzJ\varphi\right)\;,\label{eq:Num}\\
R_{\downarrow\uparrow}(\varphi)+R_{\uparrow\downarrow}(\varphi)= &  & 2\pi\left|J_{0}\left({2h}/{\omega}\right)\right|^{2}\nu\left(\left|2zJ\varphi\right|\right)\;\coth\left(\beta|zJ\varphi|\right)\nonumber \\
 &  & +2\pi\sum_{n>0}\sum_{a,b=\pm}\left|J_{n}\left({2h}/{\omega}\right)\right|^{2}\,\nu\left(an\omega+2bzJ\varphi\right)\,\coth(\beta(an\omega+2bzJ\varphi)/2)\;.\label{eq:Den}
\end{eqnarray}
\end{subequations} \end{widetext}

In case the ac drive is switched off, $h=0$, one naturally recovers
\begin{equation}
\pm\varphi=\frac{R_{\downarrow\uparrow}(\varphi)-R_{\uparrow\downarrow}(\varphi)}{R_{\downarrow\uparrow}(\varphi)+R_{\uparrow\downarrow}(\varphi)}\xrightarrow[h=0]{}\tanh\beta zJ\varphi\;,
\end{equation}
which is the familiar self-consistent condition for the Ising model
in thermal equilibrium. In this case, it is well known that there
is a second-order phase transition at the critical temperature $T_{\rmc}^{{\rm eq}}=z\left|J\right|$,
below which ferromagnetic solutions are possible for $J>0$ and anti-ferromagnetic
ones for $J<0$.

\paragraph{Non-equilibrium steady-state phase diagram.}

The self-consistency equation~(\ref{eq:Self_consistency}) together
with Eqs.~(\ref{eq:Num}) and (\ref{eq:Den}) allow us to explore
the complete mean-field phase diagram far from the equilibrium regime.
Let us first investigate the fate of the well-known second-order phase
transition in this out-of-equilibrium context. In order to access
its locus in parameter space, we expand and solve Eq.~(\ref{eq:Self_consistency})
around $\varphi=0$. Using the low-energy parametrization of the bath
hybridization $\nu(\epsilon)\stackrel[\epsilon\to0^{+}]{}{\simeq}\eta\,\epsilon^{1+s}$,
we obtain 
\begin{equation}
\pm\varphi=\frac{R_{\downarrow\uparrow}(\varphi)-R_{\uparrow\downarrow}(\varphi)}{R_{\downarrow\uparrow}(\varphi)+R_{\uparrow\downarrow}(\varphi)}=\beta zJ\varphi\;\frac{K\,|2zJ\varphi|^{s}+A}{K\,|2zJ\varphi|^{s}+B}\;,\label{eq:ratio}
\end{equation}
where 
\begin{eqnarray*}
K & \equiv & \eta\;\left|J_{0}\left({2h}/{\omega}\right)\right|^{2}\;,\\
A & \equiv & 2\,\sum_{n>0}\left|J_{n}\left({2h}/{\omega}\right)\right|^{2}\,\nu^{\,\prime}(n\omega)\;,\\
B(T) & \equiv & \beta\sum_{n>0}\left|J_{n}\left(\frac{2h}{\omega}\right)\right|^{2}\nu\left(n\omega\right)\coth\left(\frac{\beta n\omega}{2}\right).
\end{eqnarray*}
Besides the trivial solution $\varphi=0$, the self-consistent mean-field
equation~(\ref{eq:ratio}) admits non-zero solutions 
\begin{equation}
|\varphi|=\frac{1}{2\, T_{\rmc}^{{\rm eq}}}\left[\frac{B(T)}{K}\;\frac{\pm{\rm sgn}(J)\,[A/B(T)]\; T_{\rmc}^{{\rm eq}}-T}{T\mp{\rm sgn}(J)\, T_{\rmc}^{{\rm eq}}}\right]^{1/s}\!\!\!\!\!\!\!\!.\label{eq:main}
\end{equation}
Equation~(\ref{eq:main}) above is quite rich and its analysis below
will tell us about 1) the critical temperature, 2) the nature of the
ordered phase (and the stability of the non-trivial solutions), 3)
the critical exponent, and 4) the nature of the phase transition.

\begin{figure}[t]
\includegraphics[width=7cm]{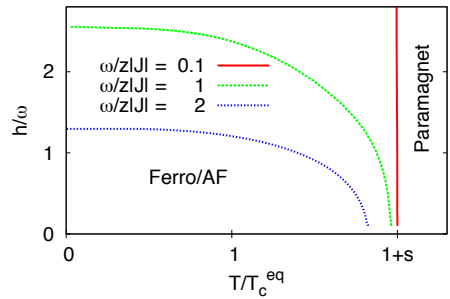} \protect\caption{{\footnotesize{}Non-equilibrium phase diagram in the drive vs temperature
plane for different values of $\omega$ given in the key. ($s=1$).}}
\end{figure}

Note that $\varphi$ in Eq.~(\ref{eq:main}) must vanish continuously
when crossing a second-order phase transition. For a bath with a sub-Ohmic
low-energy behavior, $-1\leq s<0$, this implies that the corresponding
critical temperature, $T_{\rmc}$, is identical to the equilibrium
case: $T_{\rmc}=T_{\rmc}^{{\rm eq}}$. Thereafter, unless stated otherwise,
we shall focus on baths with a super-Ohmic low-energy behavior, $s>0$.
In this case, the critical temperature is the non-trivial solution
of $T_{\rmc}=\pm{\rm sgn}(J)\,[A/B(T_{\rmc})]\; T_{\rmc}^{{\rm eq}}$.
Before solving explicitly for $T_{\rmc}$, one can already remark
that $T_{\rmc}$ must be larger than $T_{\rmc}^{{\rm eq}}$ so that
the numerator and denominator of Eq.~(\ref{eq:main}) have the same
sign for $T_{\rmc}^{{\rm eq}}<T<T_{\rmc}$, ensuring a well-defined
non-zero magnetization solution in that temperature range. Let us
now solve for $T_{\rmc}$ by considering the case when $h\ll\omega$,
for which only the $n=1$ mode contributes significantly (because
of the stronger power decay of the Bessel functions for larger $n$'s).
In this case, the critical temperature $T_{\rmc}$ is determined by
\begin{equation}
\tanh\left(\frac{\omega}{2\, T_{\rmc}}\right)=\pm{\rm sgn}(J)\;\frac{1}{2T_{\rmc}^{{\rm eq}}}\,\frac{\nu\left(\omega\right)}{\nu^{\,\prime}\left(\omega\right)}.\label{eq:Tc}
\end{equation}
Note that Eq.~(\ref{eq:Tc}) has a finite solution only if the norm
of the right-hand side is smaller than unity.

Importantly, when $\nu^{\,\prime}(\omega)>0$, the type of order is
dictated by the sign of J in the ordinary way: $J>0$ for a ferromagnet,
$J<0$ for an anti-ferromagnet. However, it is noteworthy that driving
can turn a ferromagnet into an anti-ferromagnet and vice-versa when
$\nu^{\,\prime}(\omega)<0$. The choice of sign in Eq.~(\ref{eq:Tc})
that yields a solution (phase transition) in this case is the \textit{opposite}
of the common Ising model: here when $J>0$, there is an \textit{anti-ferromagnetic}
solution, and when $J<0$, there is a \textit{ferromagnetic solution}.

Eq.~(\ref{eq:Tc}) can be solved analytically when the right-hand
side of the equation is much smaller than unity, $\nu(\omega)/|\nu'(\omega)|\ll T_{\rmc}^{{\rm eq}}$,
yielding the critical temperature 
\begin{align}
T_{\rmc}\approx T_{\rmc}^{{\rm eq}}\;\left|\omega\;{\nu^{\,\prime}\left(\omega\right)}\right|/{\nu\left(\omega\right)}\;.\label{eq:Tc2}
\end{align}
Eq.~(\ref{eq:Tc2}) transparently elucidates that by judiciously
choosing the driving frequency or engineering the bath, or both, one
can achieve a rather large critical temperatures $T_{\rmc}$, much
larger than the one for the undriven system, $T_{\rmc}^{{\rm eq}}$.
To exemplify this point, let us assume that the low-energy energy
behavior of the hybridization $\nu(\epsilon)\sim\epsilon^{1+s}$ ($s>0$)
holds up to the scale $\omega$. This yields $T_{\rmc}\approx(1+s)\, T_{\rmc}^{{\rm eq}}>T_{\rmc}^{{\rm eq}}$.
See also Fig.~(1) where we plotted the magnetization as a function
of the temperature for different drive strengths. In the temperature
range $T_{\rmc}^{{\rm eq}}<T<T_{\rmc}$, it can be seen from Eq.~(\ref{eq:main})
that the drive is responsible for a finite magnetization on the order
of $|\varphi|\sim(h/\omega)^{2/s}\omega/zJ$. In Ref.~\cite{supp},
we show the stability of this non-trivial mean-field solution below
$T_{\rmc}$. In Fig.~(2), we summarized the non-equilibrium phase
diagram in the temperature--drive plane by numerically solving for
the critical temperatures in all the regimes of $h$ and $\omega$.
Beyond the super-Ohmic case, Eq.~(\ref{eq:Tc}) suggests that one
can engineering very high critical temperatures by using the edges
of the bath spectrum to realize very large $|\nu'(\omega)|$ or by
embedding the spins in optical cavities with a finely tunable sharply
peaked spectrum.

Equation~(\ref{eq:main}) also readily provides the mean-field critical
exponent for the order parameter as function of temperature, $\beta_{T}=1/s$,
to be contrasted with the undriven case where the mean-field exponent
is $\beta_{T}^{{\rm eq}}=1/2$. This means that, even at the mean-field
level, driving changes the nature the phase transition.

Finally, Eq.~(\ref{eq:main}) predicts a diverging magnetization
at $T=T_{\rmc}^{{\rm eq}}$. Although it was derived under the assumption
that $\varphi$ is small, this suggests that the original self-consistency
Eq.~(\ref{eq:Self_consistency}) may have non-trivial solutions $\varphi\neq0$
which are not connected continuously to $\varphi=0$ and signaling
the presence of a first-order phase transition. For example, in the
case of baths with a sub-Ohmic low-energy behavior ($-1\leq s<0$),
the denominator of Eq.~(\ref{eq:Self_consistency}) given in Eq.~(\ref{eq:Den})
has $1/(\varphi-\varphi_{n})$ divergences located at every $\varphi_{n}\equiv n\omega/2zJ$
for $n=1\ldots\left\lfloor 2zJ/\omega\right\rfloor $. In turn, this
implies the presence of a collection of non-trivial solutions of the
self-consistent Eq.~(\ref{eq:Self_consistency}) close to these $\varphi_{n}$'s.
For baths with a super-Ohmic low-energy behavior, the denominator
Eq.~(\ref{eq:Den}) is well-behaved and we investigate the possibility
of a first-order phase transition by solving Eq.~(\ref{eq:Self_consistency})
numerically. In Fig.~(3), we show the non-equilibrium phase diagram
in the $T$--$\omega$ plane for a fixed $h/\omega$. Starting from
small drive frequencies, the line of second-order phase transitions
reaches a tricritical point located at $(\omega^{*}(h/\omega),T_{\rmc}^{*}=T_{\rmc}^{{\rm eq}})$
and turns into a line of first-order transitions for larger $\omega$.

\begin{figure}[t]
\includegraphics[width=7cm]{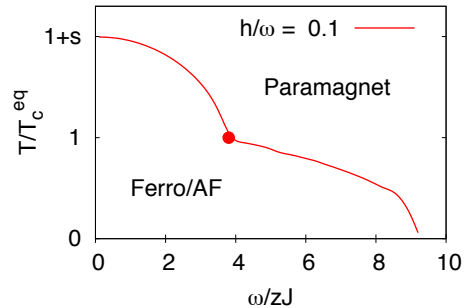} \protect\caption{{\footnotesize{}Non-equilibrium phase diagram in the temperature vs
drive frequency plane for fixed $h/\omega=0.1$. The red circle indicates
the location of the tricritical point separating a second-order line
below from a first-order line above. ($s=1$).}}
\end{figure}

\textit{Discussion.} Besides the demonstration that driven-dissipative
conditions can strongly reshape the phase diagram of the Ising model,
this study allows us to shine a new light on the fate of the universal
properties of this model and, by extension, other similar models when
driven to non-equilibrium steady states. When the drive is finite,
we have found that the critical exponents (and the critical temperature)
are strongly dependent on the details of the bath, thus loosing much
of their universality. We hope to report soon on the influence of
dimensionality (\textit{i.e.} away from the mean-field approach) on
these results by studying the one-dimensional case \textit{via} exact
methods~\cite{us}.

This work has been supported by the Rutgers CMT fellowship (G.G.),
the NSF grant DMR-115181 (C.A.), and the DOE Grant DEF-06ER46316 (C.C.).

\appendix

\section*{Supplementary Material}

\subsection*{\label{sec:stability}Stability of the mean-field solutions}

Here we check whether the non-zero mean-field solutions in Eq.~(\ref{eq:main})
are stable. We start with a Master Equation for the probabilities
$P_{\uparrow}$ and $P_{\downarrow}$ in terms of the rates ${R_{\downarrow\uparrow}}$
and ${R_{\uparrow\downarrow}}$: 
\begin{align*}
\dot{{\rm P}}_{\uparrow}= & -{R_{\uparrow\downarrow}}\,{P_{\uparrow}}+{R_{\downarrow\uparrow}}\,{P_{\downarrow}}\\
\dot{{\rm P}}_{\downarrow}= & +{R_{\uparrow\downarrow}}\,{P_{\uparrow}}-{R_{\downarrow\uparrow}}\,{P_{\downarrow}}\;.
\end{align*}
Using $P_{\uparrow}=(1\pm\varphi)/2$ and $P_{\downarrow}=(1\mp\varphi)/2$
for the ferromagnetic and anti-ferromagnetic cases, respectively,
yields 
\begin{align*}
\pm\dot{\varphi}= & \left[{R_{\downarrow\uparrow}}(\varphi)-{R_{\uparrow\downarrow}}(\varphi)\right]-\left[{R_{\downarrow\uparrow}}(\varphi)+{R_{\uparrow\downarrow}}(\varphi)\right]\;(\pm\varphi)
\end{align*}
or, equivalently, 
\begin{align*}
\dot{\varphi}= & -\left[{R_{\downarrow\uparrow}}(\varphi)+{R_{\uparrow\downarrow}}(\varphi)\right]\left\{ \varphi\mp\frac{{R_{\downarrow\uparrow}}(\varphi)-{R_{\uparrow\downarrow}}(\varphi)}{{R_{\downarrow\uparrow}}(\varphi)+{R_{\uparrow\downarrow}}(\varphi)}\;\right\} \;.
\end{align*}
The quantity in curly brackets vanishes at the stationary point, and
gives precisely the condition in Eq.~(\ref{eq:Self_consistency}).
Let $\bar{\varphi}$ be this stationary point solution. To consider
the stability of fluctuations, we expand $\varphi=\bar{\varphi}+\delta\varphi$.
The expansion of the terms in curly brackets start at order $\delta\varphi$
(because $\bar{\varphi}$ is where it vanishes); so to lowest order,
the term in square brackets does not need to be expanded. The linearized
stability equation becomes 
\begin{align*}
\dot{\delta\varphi}= & -\left[{R_{\downarrow\uparrow}}(\bar{\varphi})+{R_{\uparrow\downarrow}}(\bar{\varphi})\right][1\mp C(\bar{\varphi})]\;\delta\varphi\;,
\end{align*}
where 
\begin{align*}
C(\bar{\varphi})=\frac{d}{d\varphi}\left(\frac{{R_{\downarrow\uparrow}}(\varphi)-{R_{\uparrow\downarrow}}(\varphi)}{{R_{\downarrow\uparrow}}(\varphi)+{R_{\uparrow\downarrow}}(\varphi)}\right)\Bigg|_{\bar{\varphi}}\;.
\end{align*}
Notice that ${R_{\downarrow\uparrow}}(\bar{\varphi})+{R_{\uparrow\downarrow}}(\bar{\varphi})>0$,
so the stability of the solution rests upon whether $[1\mp C(\bar{\varphi})]>0$.

Using Eq.~(\ref{eq:ratio}), we find 
\begin{align*}
1\mp C(\bar{\varphi})= & \mp\beta zJ\,\frac{d}{d\varphi}\;\frac{K\,|2zJ\varphi|^{s}+A}{K\,|2zJ\varphi|^{s}+B}\Bigg|_{\bar{\varphi}}\\
= & \mp\beta zJ\,(A-B)\left[\frac{d}{d\varphi}\;\frac{1}{K\,|2zJ\varphi|^{s}+B}\Bigg|_{\bar{\varphi}}\right]\;.
\end{align*}
The quantity in the square bracket above is always negative. Therefore,
the sign of $1\mp C(\bar{\varphi})$ is that of $\pm{\rm sgn}(J)\,(A-B)$.
Now recall that $T_{\rmc}=\pm{\rm sgn}(J)\,[A/B(T_{\rmc})]\; T_{\rmc}^{{\rm eq}}$
is larger than $T_{\rmc}^{{\rm eq}}$ for the non-trivial magnetization
to be well defined; therefore $\pm{\rm sgn}(J)\, A>B$. Thus, $\pm{\rm sgn}(J)\,(A-B)>B[1\mp{\rm sgn}(J)]\ge0$.
Hence, we conclude that the sign of $1\mp C(\bar{\varphi})$ is positive
and the solutions we found are stable.
\end{document}